\documentclass[prb,aps,preprint]{revtex4}
\usepackage{epsfig}
\usepackage{bm}
\begin{document}

\title{A half-step in quantized conductance for low-density electrons in a quantum wire}

\author{Godfrey Gumbs$^{1,4}$}
\email{ggumbs@hunter.cuny.edu}
\author{Antonios Balassis$^{2}$}
\email{balassis@fordham.edu}
\author{Danhong Huang$^{3}$}
\email{danhong.huang@kirtland.af.mil}
\author{Sheehan Ahmed$^{2}$}
\email{ashahmed@fordham.edu}
\author{Ryan Brennan$^{2}$}
\email{rybrennan@fordham.edu}
\address{$^{1}$Department of Physics and Astronomy, Hunter College of the
City University of New York, 695 Park Avenue, New York, NY 10065, USA}
\address{ $^{2}$Physics Department, Fordham University, 441 East Fordham Road,
       Bronx, NY 10458, USA}
\address{$^3$Air Force Research Laboratory, Space Vehicles
Directorate,\\
Kirtland Air Force Base, NM 87117, USA}
\address{$^4$Donostia International Physics Center (DIPC),
P. de Manuel Lardizabal, 4, 20018 San Sebasti\'an,
Basque Country, Spain }

\date{\today}

\begin{abstract}
We investigated the effect due to perpendicular magnetic  field  on quantum wires
 where spin-orbit interaction (SOI) of electrons is not neglected.  Based on the
 calculated energy dispersion, the nonlinear ballistic conductance ($G$) and
 electron-diffusion thermoelectric power ($S_d$) are calculated as functions of electron
 density, temperature and applied bias voltage. A low-temperature half-step feature
 in $G$, which was observed experimentally by Quay {\em et al.\/} [see Nature
 Physics {\bf 6}, 336 (2010)], as well as a new peak in $S_d$ are reproduced
here in the low density regime. These phenomena are related to the
occurrence of the Zeeman splitting and SOI induced saddle point
in the band structure, where the channel chemical potential lies within
an anticrossing gap between the saddle point of the lower subband and the
bottom of the upper subband. Additionally, side peaks in $G$ far away from the
zero bias for the nonlinear transport, as well as a quadratic bias-voltage
dependence of $G$ near zero voltage, are predicted and discussed.
\end{abstract}

\pacs{73.23.Ad, 75.70.Tj, 81.07.Gf}
\maketitle

\section{Introduction}
\label{sec1}

Quasi-one-dimensional (quasi-1D) electron systems (also called
quantum wires) have attracted considerable attention both theoretically\,\cite{sarma1,pollak,book}
and experimentally\,\cite{hansen,heitmann} since their experimental realization in
1986.\,\cite{thornton} When an external magnetic field is applied in a
direction perpendicular to a quantum wire, Landau quantization in
such a system combines with lateral confinement, enhancing the
quantum-wire confinement potential as well as the effective mass of
electrons along the wire direction at the same time.\,\cite{stormer,gumbs1}
Several studies were reported for such  systems, including
magneto-optical properties\,\cite{heitmann,horing} and magneto-transport properties\,\cite{weiss,chao,lyo1,lyo2,lyo3,lyo4} in the diffusive regime. Moreover,
the ballistic transport properties\,\cite{fischer,mourokh,pepper} of
quantum wires were also explored under a tilted magnetic field.
\medskip

On the other hand, the effect of Rashba spin-orbit interaction
(SOI) on the ballistic and diffusive electron transport properties in quantum
wires has also been investigated theoretically and
experimentally in the absence of a magnetic
field.\,\cite{moroz1,moroz2,pershin,gumbs2,gumbs} Compared with the
case when SOI is neglected,\,\cite{lyo5} the quantum-wire conductance quantization in the
presence of SOI is modified significantly at low temperatures.\,\cite{gumbs}
Although the SOI-induced gap has been observed experimentally,\,\cite{quay}
there are only a few theoretical studies dealing with  ballistic electron
transport and thermoelectric properties in parabolically-confined quantum
wires with both SOI and a magnetic field.\,\cite{debald,zhang,lu}
\medskip

In this paper, we include the effects due to the edges of  a realistic
quantum wire through sharp and high potential walls
forming the boundaries, instead of assuming a parabolic potential
for the lateral confinement of the quantum-wire . Extending our previous
study for zero magnetic field in the linear response regime,\,\cite{gumbs}
we calculate the bias-voltage dependent nonlinear ballistic conductance
and electron-diffusion thermoelectric power for this
quasi-one-dimensional system with a channel length less
than the electron mean free path, where we have included the
effect on the SOI of a finite magnetic field perpendicular to the
quantum wire as well as the Zeeman effect on the electron spin states.
A half-step feature in the quantized conductance is obtained for low
electron densities and low temperatures.
\medskip

The rest of this paper is arranged as follows. In
Sec.\,\ref{sec2}, we present our model for quantum wires with SOI under a
perpendicular magnetic field. Numerical results and discussion of the energy bands,
conductance and diffusion thermoelectric power as functions of
electron density, temperature and bias voltage are presented in Sec.\,\ref{sec3}.
We summarize our results in Sec.\,\ref{sec4}.

\section{The Model}
\label{sec2}

In the presence of a magnetic field ${\bf B}$ (in the $z$ direction)
perpendicular to a quantum wire which is confined in the $x$
direction such that $|x|\leq{\cal W}/2$ and extends along the $y$
direction, the spin-coupled Schr\"odinger equations for the electron
wave-function spinor,
$[\psi^{\uparrow}_{k_y}(x),\,\psi^{\downarrow}_{k_y}(x)]^T$, with
the Landau gauge are written as\,\cite{gumbs}

\[
-\frac{\hbar^2}{2m^\ast}\left[\frac{d^2}{dx^2}-\left(k_y+\frac{x}{\ell^2_c}\right)^2\right]
\psi^{\uparrow}_{k_y}(x)+\alpha\left[\frac{d}{dx}+\left(k_y
+\frac{x}{\ell^2_c}\right)\right]\psi^{\downarrow}_{k_y}(x)
\]
\begin{equation}
-\beta\,\frac{{\cal
W}}{\ell_0}\left(k_y+\frac{x}{\ell^2_c}\right){\cal
F}(x)\,\psi^{\uparrow}_{k_y}(x)
=\left(\varepsilon-\frac{1}{2}\,g^\ast\hbar\omega_c\right)\,\psi^{\uparrow}_{k_y}(x)\
, \label{e1}
\end{equation}

\[
-\frac{\hbar^2}{2m^\ast}\left[\frac{d^2}{dx^2}-\left(k_y+\frac{x}{\ell^2_c}\right)^2\right]
\psi^{\downarrow}_{k_y}(x)-\alpha\left[\frac{d}{dx}-\left(k_y+
\frac{x}{\ell^2_c}\right)\right]\psi^{\uparrow}_{k_y}(x)
\]
\begin{equation}
+\beta\,\frac{{\cal
W}}{\ell_0}\left(k_y+\frac{x}{\ell^2_c}\right){\cal
F}(x)\,\psi^{\downarrow}_{k_y}(x)
=\left(\varepsilon+\frac{1}{2}\,g^\ast\hbar\omega_c\right)\,\psi^{\downarrow}_{k_y}(x)\ ,
\label{e2}
\end{equation}
where $\beta$ and $\alpha$ are the SOI parameters in the transverse $x$
  and $z$ directions, respectively. Also,
  $\varepsilon$ is the eigenenergy of the electrons, the
symbols $\uparrow$ and $\downarrow$ denote the up and down spin
states, $k_y$ is the electron wave number along the wire, $m^\ast$
is the effective mass, $\omega_c=eB/m^\ast$ is the cyclotron
frequency, $\ell_c=\sqrt{\hbar/eB}$ is the magnetic length, and
$g^\ast$ is the effective Lande spin $g-$factor. Additionally, the
quantum confinement factor\,\cite{gumbs} introduced in the above
equations for a non-parabolic confinement is given by

\begin{equation}
{\cal F}(x)=\exp\left[-\frac{(x+{\cal
W}/2)^2}{2\ell_0^2}\right]-\exp\left[-\frac{(x-{\cal
W}/2)^2}{2\ell_0^2}\right]\ ,
\label{e3}
\end{equation}
where $\ell_0$ describes the steepness of the confinement potential
at the two edges of the wire. Compared with the case of $B=0$,\,\cite{gumbs} we
see that $k_y$ is replaced by $(k_y+x/\ell^2_c)$, as well as the appearance of a
Zeeman splitting term, in Eqs.\,(\ref{e1}) and (\ref{e2}) when $B\neq0$.
\medskip

For ballistic electrons in a quantum wire, the charge ($n=0$) and
the heat ($n=1$) currents are calculated as\,\cite{lyo5}

\begin{eqnarray}
{\cal Q}^{(n)}(V_b)&=& (-e)^{1-n}\,\frac{1}{2\pi}\,\sum_j\,\int\limits_{-\infty}^\infty dk_y\,\left(\varepsilon_{jk_y}-\mu_{\rm c}\right)\,v_{jk_y}
\nonumber\\
&\times &  \left[\theta(-v_{jk_y})\,f_{jk_y}^{({\rm
R})}\left(1-f_{jk_y}^{({\rm L})}\right)
+\theta(v_{jk_y})\,f_{jk_y}^{({\rm L})}\left(1-f_{jk_y}^{({\rm
R})}\right)\right]\ .
 \label{e4}
\end{eqnarray}
In this notation, $j=1,\,2,\,\cdots$ is a composite index labeling the
spin-resolved subbands,
$v_{jk_y}=(1/\hbar)\,d\varepsilon_{jk_y}/dk_y$ is the group velocity
of electrons in the $j$th subband, $\theta(x)$ is the unit step
function, $\mu_{\rm c}$ is the chemical potential of the quantum wire,
$f^{(s)}_{jk_y}=\{1+\exp[(\varepsilon_{jk_y}-\mu_s)/k_BT]\}^{-1}$,
$T$ is the temperature of the system, $s={\rm R},\,{\rm L}$ denotes
the right and left electrodes connected to two ends of a quantum
wire in the $y$ direction, $\mu_{\rm L}=\mu_{\rm c}-eV_b/2$ and
$\mu_{\rm R}=\mu_{\rm c}+eV_b/2$ represent the chemical potentials
of the left and right electrodes, satisfying $\mu_{\rm R}-\mu_{\rm
L}=eV_b$ with $V_b$ being the external bias voltage. For fixed
chosen  linear electron density $n_{\rm 1D}$ and $T$, $\mu_{\rm
c}$ can be calculated using the following constraint equation

\begin{equation}
\sum_j\left\{\ \int\limits_{-\infty}^\infty\,
\frac{dk_y}{\exp\left[(\varepsilon_{jk_y}-\mu_{\rm
c})/k_BT\right]+1}\right\}-2\pi n_{\rm 1D}=0\ . \label{e7}
\end{equation}
Once the ${\cal Q}^{(n)}(V_b)$ in Eq.\,(\ref{e4}) has been calculated,
the bias-dependent ballistic conductance $G(V_b)$ of the quantum wire is
simply given by

\begin{equation}
G(V_b)=\frac{{\cal Q}^{(0)}(V_b)}{V_b}\ ,
\label{e5}
\end{equation}
and the diffusion thermoelectric power $S_{\rm d}(V_b)$ may be calculated
according to

\begin{equation}
S_{\rm d}(V_b)=\frac{1}{T}\left[\frac{{\cal Q}^{(1)}(V_b)}{{\cal Q}^{(0)}(V_b)}\right]\ .
\label{e6}
\end{equation}
We now employ these quantities to carry out our numerical calculations.

\section{Numerical Results}
\label{sec3}

In our calculations, we define two spin-related
characteristic lengths, namely, $\ell_\alpha=\hbar^2/(2m^\ast\alpha)$ and
$\ell_\beta=\hbar^2/(2m^\ast\beta)$. Consequently, four corresponding
dimensionless quantities related to  spin-orbit coupling, quantum
confinement factor and magnetic-field effect are given by:
$\tau_\alpha={\cal W}/\ell_\alpha\propto\alpha$, $\tau_\beta={\cal
W}/\ell_\beta\propto\beta$, $\tau_0={\cal W}/\ell_0$ and
$\tau_c={\cal W}/\ell_c\propto\sqrt{B}$. The scaled electron kinetic
energy is $\varepsilon_{jk_y}/E_0$ with
$E_0=\pi^2\hbar^2/2m^\ast{\cal W}^2$, the scaled electron wave
number is $k_y{\cal W}$, and the scaled position variable is
$x/{\cal W}$. Parameters in our calculations were chosen as
$m^\ast=0.067\,m_0$ (as in GaAs) with $m_0$ being the free-electron mass,
$g^\ast=-0.44$ due to exchange interaction, $\tau_\beta=10$, ${\cal
W}=568.7$\,\AA\ and $\tau_0=10^3$. Other parameters, such as
$\tau_\alpha$, $B$, $T$, $V_b$ and $n_{\rm 1D}$, will be given in
the figure captions.
\medskip

\begin{figure}[p]
\begin{center}
\epsfig{file=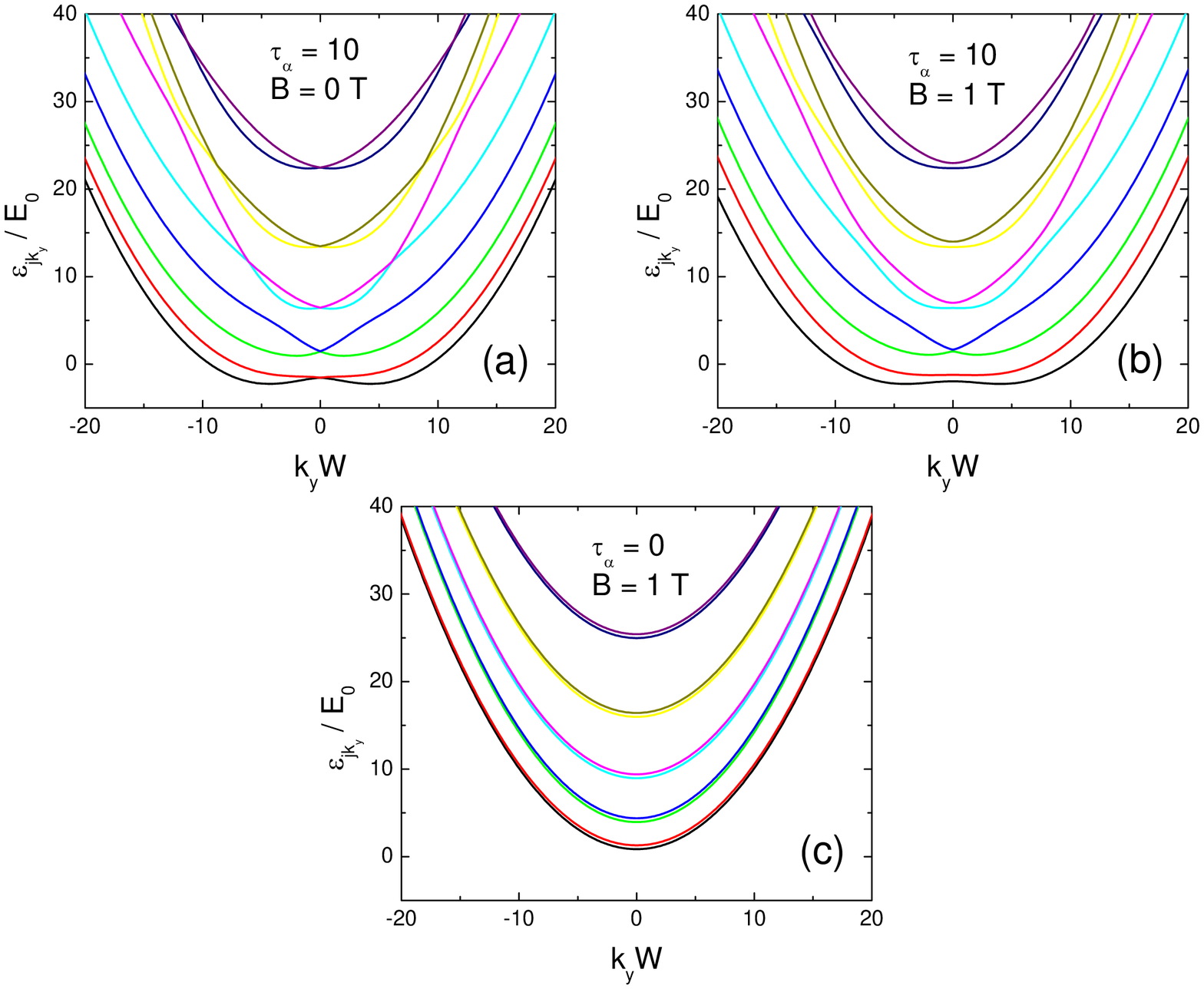,width=6in} \caption{(Color online)
The calculated eigenenergies $\varepsilon_{jk_y}/E_0$ as
functions of $k_y{\cal W}$ for the three cases considered: (a)
$\tau_\alpha=10$ and $B=0$\,T; (b) $\tau_\alpha=10$ and $B=1$\,T;
(c) $\tau_\alpha=0$ and $B=1$\,T. The Zeeman splitting in (c) can
be resolved for $B=1$\,T. The degeneracy of $\uparrow$ and $\downarrow$ spin states at $k_y=0$
are split by the Zeeman effect in (b) to form an anticrossing gap. }
\label{f1}
\end{center}
\end{figure}

Figure\ \ref{f1} shows the calculated eigenenergy
$\varepsilon_{jk_y}/E_0$ as a function of $k_y{\cal W}$ for three
different cases, i.e., (a) $\tau_\alpha=10$, $B=0$\,T; (b)
$\tau_\alpha=10$, $B=1$\,T; (c) $\tau_\alpha=0$, $B=1$\,T. From
Fig.\,\ref{f1}(a), we find that the spin-orbit interaction under
$\alpha\neq 0$ couples spin-up and spin-down electron states in the
absence of a $B$ field. Here, the energy dispersion curves of the
spin-up and spin-down states are shifted relative to each other
with respect to $k_y=0$. In the presence of a $B$ field, as shown in
Fig.\,\ref{f1}(b), many anticrossing gaps are created due to both Zeeman splitting and SOI, the effective
mass of electrons along the wire is enhanced, and new saddle points
are formed.\,\cite{lyo2} After the spin-orbit interaction is turned
off, we simply see a set of Zeeman split parabolas, as demonstrated
by Fig.\,\ref{f1}(c). When $B\rightarrow\infty$, these subbands will
convert into discrete Landau levels (with infinite effective mass
along the $y$ direction).
\medskip

\begin{figure}[p]
\begin{center}
\epsfig{file=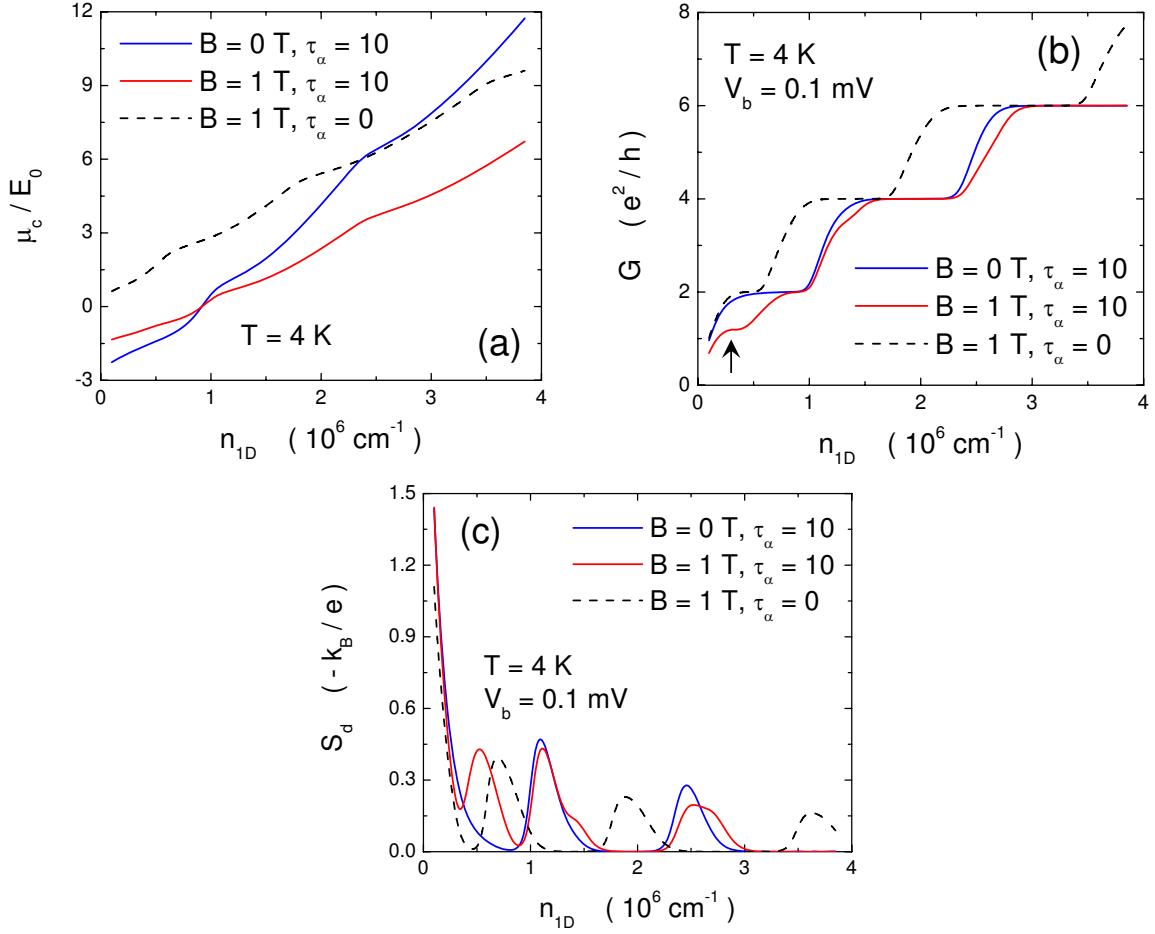,width=6in}
\caption{(Color online) (a) Chemical potential $\mu_{\rm c}$
(in units of $E_0$) for a quantum wire, (b) ballistic conductance $G$
(in units of $(e^2/h)$) and (c) diffusion thermoelectric power $S_d$ (in units of
$(-k_B/e)$) as functions of the linear density $n_{\rm 1D}$ for the three
cases: $\tau_\alpha=10$, $B=0$\,T (blue solid curves);
$\tau_\alpha=10$, $B=1$\,T (red solid curves); and $\tau_\alpha=0$,
$B=1$\,T (black dashed curves), where $T=4$\,K and $V_b=0.1$\,meV.
The black arrow in (b) indicates a half-step in quantized conductance $G$.}
\label{f2}
\end{center}
\end{figure}

The results for the quantum-wire chemical potential $\mu_{\rm
c}/E_0$, the ballistic conductance $G$ [in units of $(e^2/h)$], and
the diffusion thermoelectric power $S_{\rm d}$ [in units of
$(-k_B/e)$] are compared in Figs.\,\ref{f2}(a),\,(b) and (c),
respectively, for the three cases discussed in Fig.\,\ref{f1}. Here,
$T=4$\,K and a small bias voltage $V_b=0.1$\,meV is used for the linear-response regime. From
Fig.\,\ref{f2}(a) we see that  when $\alpha\neq 0$, $\mu_{\rm c}$ is
reduced at high electron densities but enhanced at low densities by the $B$ field
due to the formed anticrossing gap and increased electron effective mass.
Additionally, there exists a series of kinks in $\mu_{\rm c}$ as a
function of $n_{\rm 1D}$, which is related to the onset of a new
subband population. By turning off the SOI,
$\mu_{\rm c}$ is increased for finite values of $B$ and the kinks in
$\mu_{\rm c}$ are shifted at the same time due to the suppression of
the flattened region around $k_y=0$ where the density of states is expected
to be high. This series of kinks observed in Fig.\,\ref{f2}(a)
leads to the occurrence of quantum steps for $G$ in
Fig.\,\ref{f2}(b). Furthermore, each quantum step in $G$ relates to
a peak in $S_{\rm d}$, as presented in Fig.\,\ref{f2}(c). It is
interesting to note that the application of the $B$ field has a very
profound effect on the spin-orbit interaction of electrons at low
densities, where a new step (indicated by a black arrow) in (b) and a new  peak in (c) are seen
in correspondence to $\mu_{\rm c}$ lying inside the lowest
anticrossing gap. Similar half-step features were observed
experimentally.\,\cite{quay}
\medskip

\begin{figure}[p]
\begin{center}
\epsfig{file=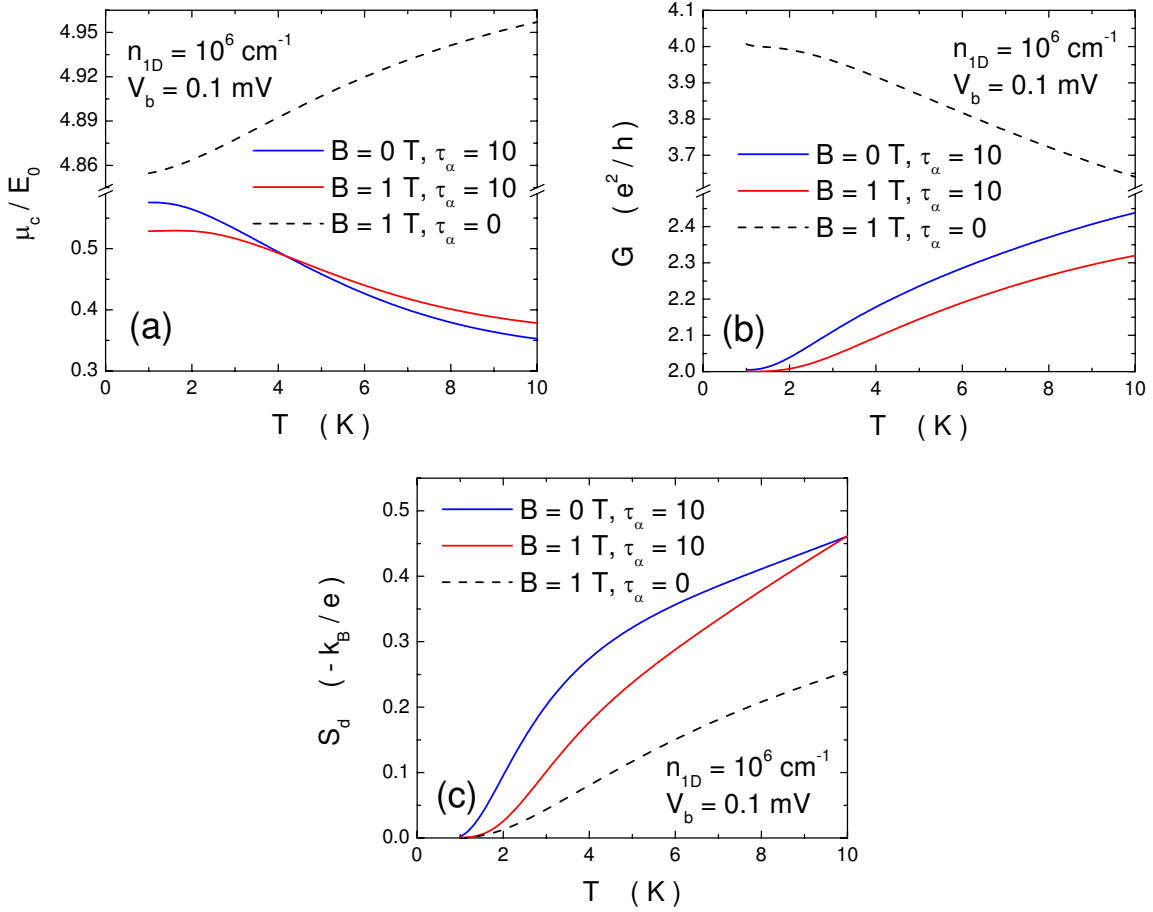,width=6in}
\caption{(Color online) Calculated (a) $\mu_{\rm c}$, (b) $G$  and   (c) $S_d$  as
functions of temperature $T$ for the three cases: $\tau_\alpha=10$,
$B=0$\,T (blue solid curves); $\tau_\alpha=10$, $B=1$\,T (red
solid curves); and $\tau_\alpha=0$, $B=1$\,T (black dashed
curves), where $n_{\rm 1D}=10^6$\,cm$^{-1}$ and $V_b=0.1$\,meV. }
\label{f3}
\end{center}
\end{figure}

Comparisons for $\mu_{\rm c}$, $G$ and $S_{\rm d}$ as a function
of temperature $T$ for  the three cases discussed in Fig.\,\ref{f1} are
presented in Figs.\,\ref{f3}(a),\,(b) and (c), respectively, where
we chose $n_{\rm 1D}=10^6$\,cm$^{-1}$ and $V_b=0.1$\,meV. With a finite
SOI, the $B$ field reduces (increases) $\mu_{\rm c}$ in Fig.\,\ref{f3}(a)
for $T<4$\,K ($T>4$\,K). For $\alpha=0$,
$\mu_{\rm c}$ is significantly enhanced. Meanwhile, $\mu_{\rm c}$, which
lies within the gap of the second and third parabolas, increases
with $T$ in contrast to the case when $\alpha\neq 0$ where a
decreasing $\mu_{\rm c}$ with $T$ is obtained due to the fact that
$\mu_{\rm c}$ stays very close to the two minima of the third
subband in Figs.\,\ref{f1}(a) and (b). For nonzero SOI,
 $G$ increases with $T$ in Fig.\,\ref{f3}(b) as a result of
the gradual thermal population of higher subbands slightly above
$\mu_{\rm c}$. On the other hand, $G$ is decreased when $T$ is
increased in the absence of spin-orbit coupling since $G\propto[-\partial
f^{(s)}_{jk_y}/\partial\varepsilon_{jk_y}]$ whose peak strength is
reduced by increasing $T$ provided $\mu_{\rm c}$ lies within a gap.
In addition, as can be seen from Fig.\,\ref{f3}(c), $S_{\rm d}$
monotonically increases with $T$, and becomes linearly
proportional to $T$ above $4$\,K.\,\cite{lyo3} However, it becomes
exponentially small for $T\ll 1$\,K.\,\cite{lyo3} For $\alpha=0$,
$S_{\rm d}$ becomes much smaller than in  the previous two cases due
to a larger value of $G$.
\medskip

\begin{figure}[p]
\begin{center}
\epsfig{file=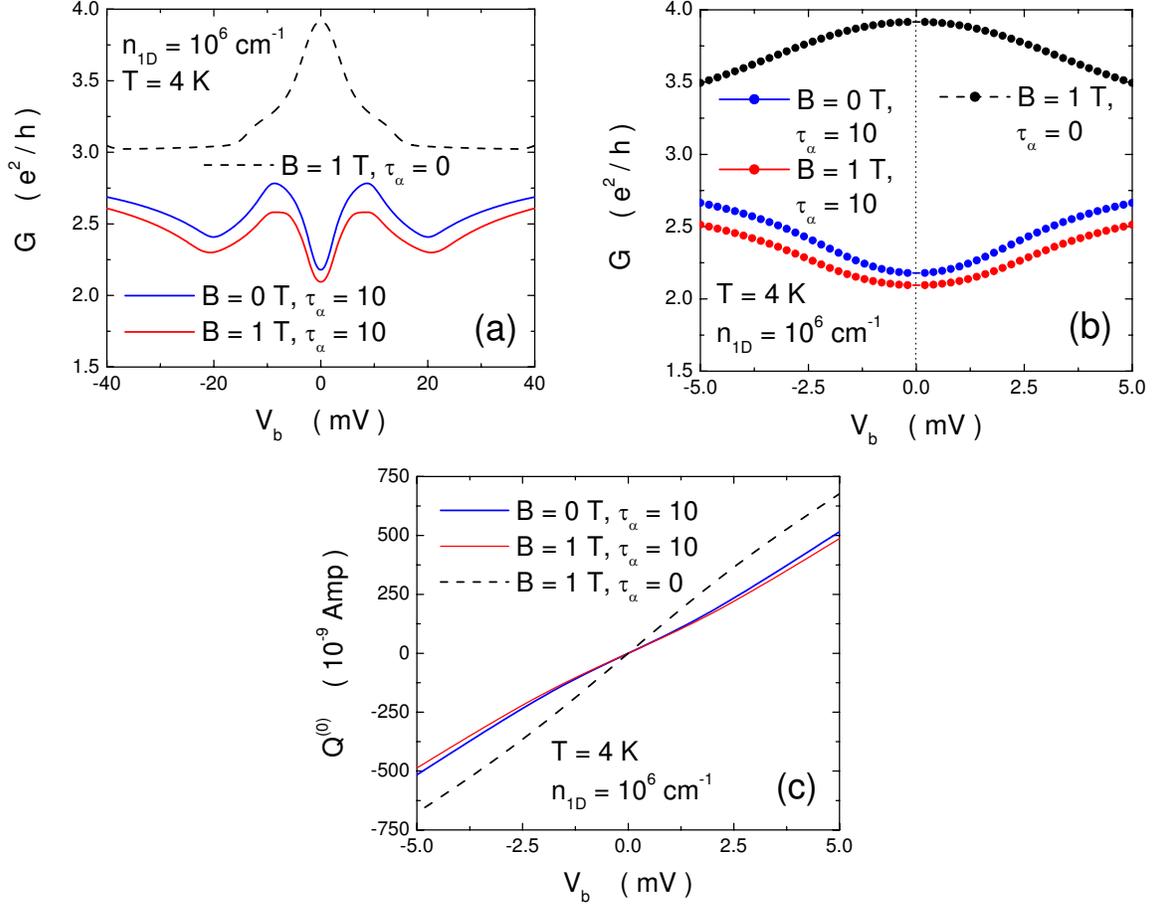,width=6in}
\caption{(Color online) Calculated (a) $G$, (b) its blow-out view and (c) charge
current ${\cal Q}^{(0)}$ as functions of the bias voltage $V_b$
for $\tau_\alpha=10$, $B=0$\,T (blue solid curves),
$\tau_\alpha=10$, $B=1$\,T (red solid curves), and
$\tau_\alpha=0$, $B=1$\,T (black dashed curves). Here, $T=4$\,K
and $n_{\rm 1D}=10^6$\,cm$^{-1}$.  The vertical dotted line in (b)
indicates the position for $V_b=0$.}
\label{f4}
\end{center}
\end{figure}

We present in Fig.\,\ref{f4}  comparisons between the bias voltage
($V_b$) dependence of $G$ in (a), the low-voltage blow-out view of
Fig.\,\ref{f4}(a) in (b) and the charge current ${\cal Q}^{(0)}$ in
(c) for the three cases considered in Fig.\,\ref{f1}, where $n_{\rm
1D}=10^6$\,cm$^{-1}$ and $T=4$\,K. When $\tau_\alpha=10$ in
Fig.\,\ref{f4}(a), $G$ initially increases with $|V_b|$ from
$2e^2/h$ in a symmetric way, which is followed by a decrease before
it increases again. The $B$ field does not alter the shape of
 the $G-V_b$ curve except for a small reduction of $G$. On the other
hand, $G$ decreases with $|V_b|$ from $4e^2/h$ and eventually
saturates at $3e^2/h$ for $\tau_\alpha=0$. These dramatically
distinct behaviors in $G$ are attributed to different positions of
the chemical potentials in the two electrodes relative to the bottom of
subbands as $V_b\rightarrow 0$, as explained in Fig.\,\ref{f3}. In
order to see the behavior of $G$  in the low-voltage regime, we display in
Fig.\,\ref{f4}(b) the blow-out view of Fig.\,\ref{f4}(a), where a
quadratic dependence of $G$ on $V_b$ can be identified. Figure\
\ref{f4}(c) exhibits a different dependence of ${\cal Q}^{(0)}$ on
$V_b$. For the cases with $\tau_\alpha=10$, the leading term in the
charge current is ${\cal Q}^{(0)}\propto V_b\,[-\partial
f^{(s)}_{jk_y}/\partial\varepsilon_{jk_y}]$ for $|V_b|\leq
1.5$\,meV. However, for $|V_b|\geq 2.5$\,meV in Fig.\,\ref{f4}(c), we
find that the leading term changes to ${\cal Q}^{(0)}\propto
\,V^3_b[-\partial^3 f^{(s)}_{jk_y}/\partial\varepsilon^3_{jk_y}]$.
This explains the observations in Fig.\,\ref{f4}(b), i.e.,
$G=2e^2/h$ for $V_b\rightarrow 0$ (linear response) and $G\propto
V_b^2$ (nonlinear response) for large values of $|V_b|$. As
$\tau_\alpha=0$, the slope in the linear-response regime is greatly
enhanced because of a larger value of the Fermi velocity (large
$\mu_{\rm c}$), leading to a quadratic decrease of $G$ with $V_b$ in
Fig.\,\ref{f4}(b) as $|V_b|$ increases.
\medskip

\begin{figure}[p]
\begin{center}
\epsfig{file=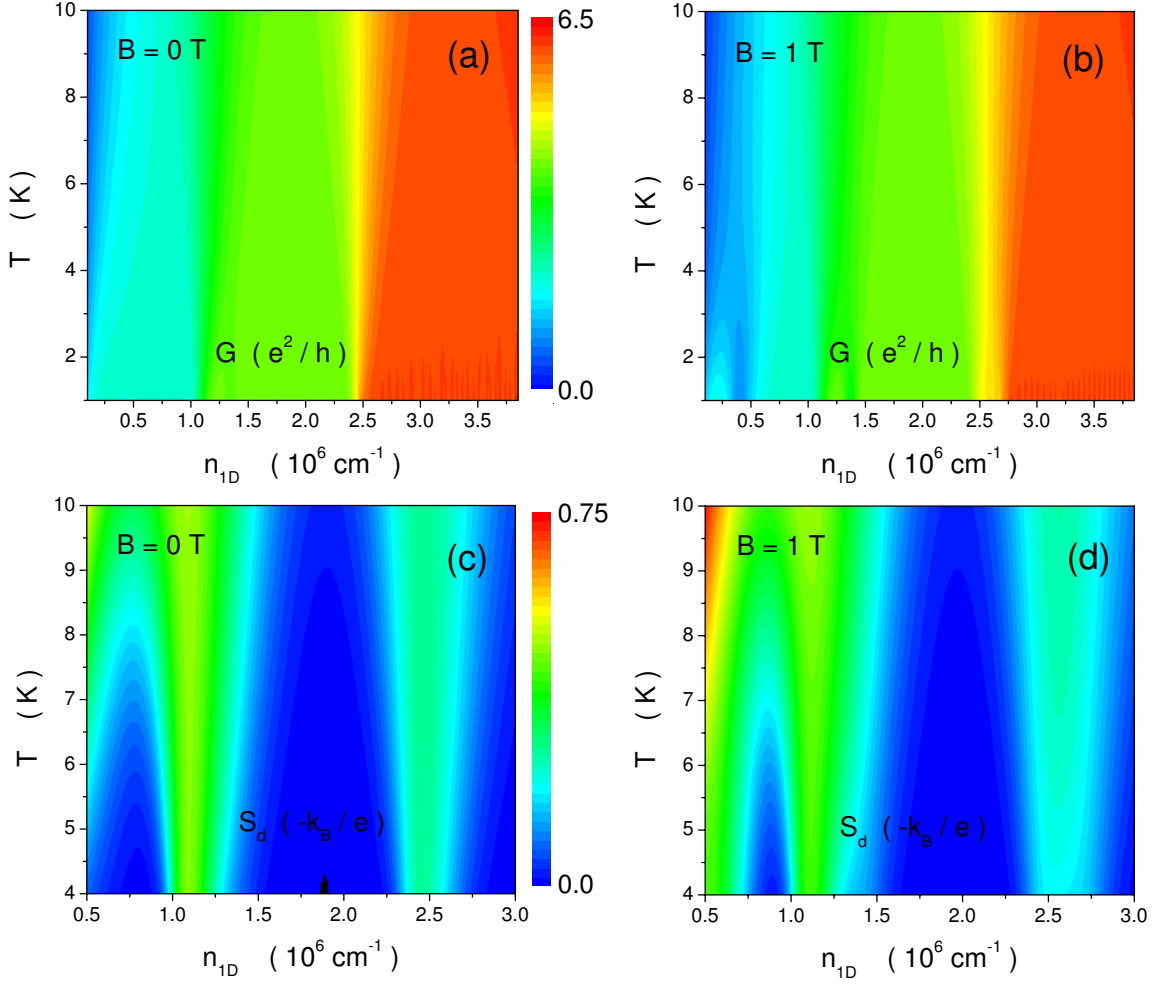,width=6in}
\caption{(Color online) Contour plots of $G$ [in (a) and (b)] and $S_d$
[in (c) and (d)] as functions of $n_{\rm 1D}$ (horizontal axis) and $T$ (vertical axis) for
$B=0$\,T, in  ((a) and (c)) and $B=1$\,T ((b) and (d)). Here,
$V_b=0.1$\,meV and $\tau_\alpha=10$. The color scale of $G$ in (a)
and (b) varies from $0$ (blue) to $6.5$ (red), whereas that of $S_d$ in
(c) and (d) varies from $0$ (blue) to $0.75$ (red).}
\label{f5}
\end{center}
\end{figure}

In Fig.\,\ref{f5} we display two-dimensional (2D) contour plots of
$G(n_{\rm 1D},\,T)$ in (a) and (b) as well as 2D contour plots
of $S_{\rm d}(n_{\rm 1D},\,T)$ in (c) and (d) when  $B=0$ [(a) and (c)]
and when $B=1$\,T [(b) and (d)]. We chose $V_b=0.1$\,meV and
$\tau_\alpha=10$. In general, for both  $B=0$ and $B=1$\,T, we
clearly see quantized steps in $G$ and peaks in $S_{\rm d}$ with
increasing $n_{\rm 1D}$ at any value of $T$ in this figure. The step
and peak positions are almost independent of $T$. However, in the low-density
regime when $n_{\rm 1D}\leq 0.5\times 10^6$\,cm$^{-1}$,  the
presence of the $B$ field modifies the spin-orbit coupling and
dramatically alters the behaviors of $G$ and $S_{\rm d}$ as
functions of both $n_{\rm 1D}$ and $T$.
\medskip

\begin{figure}[p]
\begin{center}
\epsfig{file=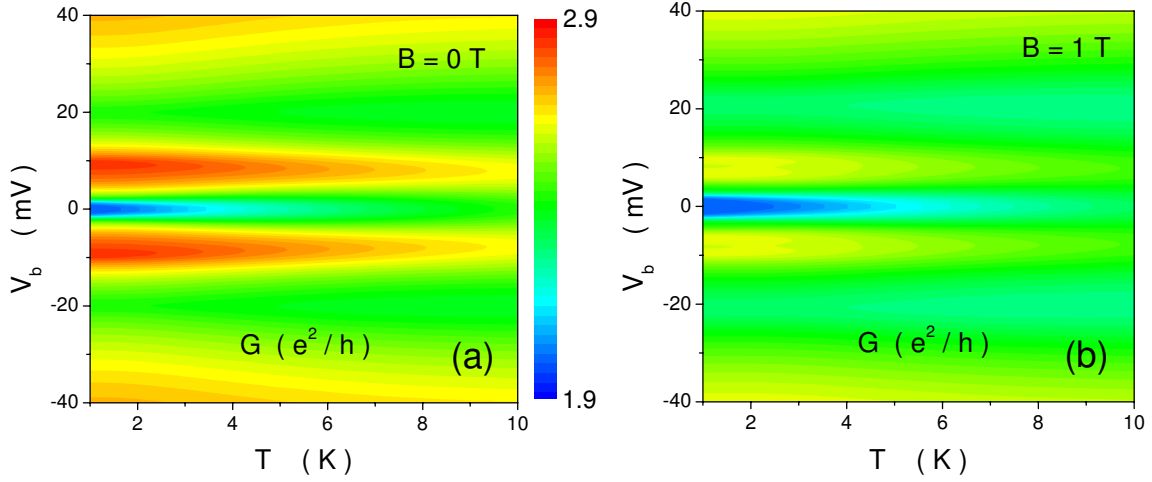,width=6in}
\caption{(Color online) Contour plots of $G$ as functions of $T$
(horizontal axis) and $V_b$ (vertical axis) for (a) $B=0$\,T and (b) $B=1$\,T.
We chose $n_{\rm 1D}=10^6$\,cm$^{-1}$ and $\tau_\alpha=10$. The color scale
for $G$ in (a) and (b) varies from $1.9$ (blue) to $2.9$ (red).}
\label{f6}
\end{center}
\end{figure}

We compare the 2D contour plots of $G(T,\,V_b)$ in Fig.\,\ref{f6} at
$B=0$ in (a) and $B=1$\,T in (b), for chosen $n_{\rm 1D}=10^6$\,cm$^{-1}$
and $\tau_\alpha=10$. It is easy to see that the two strong peaks in $G$
on either side of $V_b=0$ are greatly suppressed by the magnetic
field, which is further accompanied by an expanded dip around
$V_b=0$. In addition, the peaks in $G$ decrease with $T$.
\medskip

\begin{figure}[p]
\begin{center}
\epsfig{file=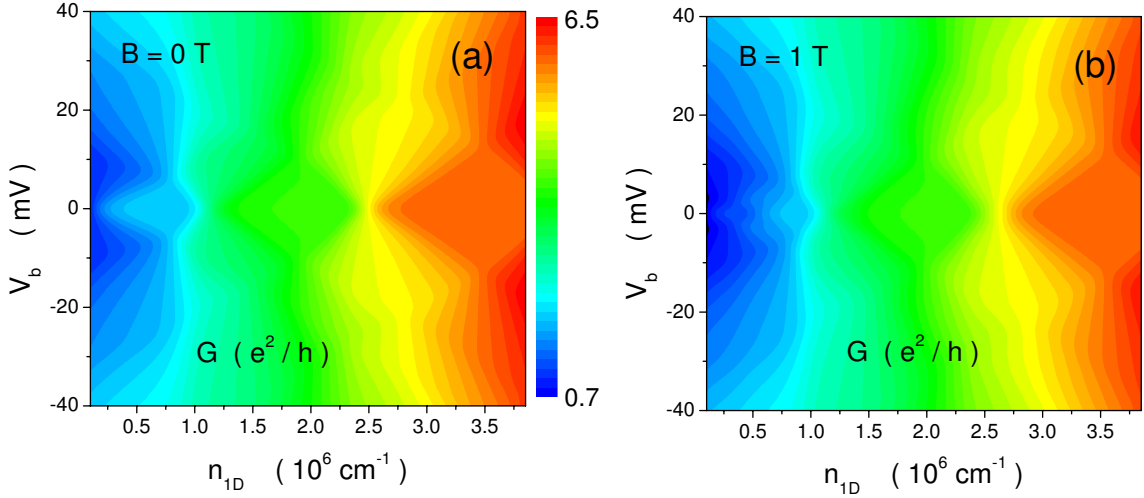,width=6in} \caption{(Color online)
Contour plots of $G$ as functions of $n_{\rm 1D}$ (transverse) and
$V_b$ (vertical) for $B=0$\,T [in (a)] and $B=1$\,T [in (b)]. Here,
$T=4$\,K and $\tau_\alpha=10$. The color scale of $G$ in (a) and (b)
is from $0.7$ (blue) to $6.5$ (red).} \label{f7}
\end{center}
\end{figure}

For completeness, we present in Fig.\,\ref{f7} the 2D contour plots of
$G(n_{\rm 1D},\,V_b)$ for $B=0$ and $B=1$\,T  in (a)
and (b), respectively, at $T=4$\,K when $\tau_\alpha=10$. It is clear that the
quantized steps in $G$ become most observable close to $V_b=0$. The
$B$ field does not significantly affect $G$ in the
high-density regime ($n_{\rm 1D}\geq 2.5\times 10^6$\,cm$^{-1}$). In
the intermediate-density regime ($1.2\times 10^6$\,cm$^{-1}\leq
n_{\rm 1D}\leq 2.5\times 10^6$\,cm$^{-1}$), $G$ only weakly depends
on $|V_b|$ and shows no $B$-field effect. In the low-density regime
$(n_{\rm 1D}\leq 1.2\times 10^6$\,cm$^{-1}$), however, the $B$ field changes $G$
substantially as functions of both $n_{\rm 1D}$ and $|V_b|$. The dip
around $V_b=0$ evolves into a peak and vice versa with increasing
$n_{\rm 1D}$ due to the variation of  the relative position of $\mu_{\rm c}$  with respect
to the bottom of the subband.

\section{Summary and Concluding Remarks}
\label{sec4}

In summary, we have investigated the combined effect of SOI and  an external
magnetic field which is applied perpendicularly to a quantum wire. The
quantities calculated are the  the quasi-one-dimensional energy
subbands,  the nonlinear ballistic conductance ($G$) and
 electron-diffusion thermoelectric power ($S_d$)  as functions of electron
 density, temperature and applied bias voltage.
    An important outcome of our numerical calculations is the    observation of a half-step
in the low-temperature quantized conductance at low electron densities. We have
 accounted for this feature as being due to the occurrence of a saddle point in the
 subband energy dispersion relation.  The saddle point is due to the
combined action of  the applied magnetic field and spin-orbit coupling in the wire.
 A similar low-temperature half-step feature in $G$  was observed experimentally
 by Quay {\em et al.\/} \cite{quay}.   A new peak in $S_d$ is produced in the
 low density regime, and is also  related to the
occurrence of the magnetic field and SOI induced saddle point
in the band structure, where the channel chemical potential lies within
an anticrossing gap between the saddle point of the lower subband and the
bottom of the upper subband.  Additionally, the present nonlinear transport theory
has predicted side peaks in the quantized conductance as a function of
bias voltage away from zero and a quadratic bias dependence near zero voltage.
This extension of our previous work \cite{gumbs} clearly demonstrates
that the application of magnetic field leads to observable differences in the ballistic transport
properties of a quantum wire structure with SOI playing a role.

\begin{acknowledgments}
This research was supported by  contract \# FA 9453-11-01-0263 of AFRL.
DH would like to thank the Air Force Office of Scientific Research
(AFOSR) for its support.
\end{acknowledgments}

\end{document}